\documentclass[11pt]{article}
\usepackage{moriond,epsfig}

\def\be{\begin{equation}}
\def\ee{\end{equation}}
\def\bea{\begin{eqnarray}}
\def\eea{\end{eqnarray}}

\def\eunit{ergs~s$^{-1}$~cm$^{-2}$}
\begin{document}
\vspace*{4cm} \title{Galaxy groups and low mass clusters at $z<0.6$: A
perspective from the XMM Large Scale Structure survey}

\author{ J.P. Willis$^1$, F. Pacaud$^2$, M. Pierre$^2$}

\address{$^1$Department of Physics and Astronomy, University of
Victoria,\\ Victoria, V8P 1A1, Canada.\\ $^2$CEA, Saclay, Service
d'Astrophysique, F-91191, Gif-sur-Yvette, France.}

\maketitle\abstracts{ Galaxy groups and low-mass clusters provide
important laboratories in which to study X-ray gas physics and the
interplay between galaxy evolution and environmental effects. The
X-ray Multi-Mirror (XMM) Large Scale Structure (LSS) survey has
currently imaged 5 deg$^2$ to a nominal extended source flux limit of
order $10^{-14}$ \eunit\ and is dominated numerically by low-mass
groups and clusters at redshifts $0.2 < z < 0.6$. We discuss the
generation of the XMM-LSS cluster sample, initial results on the
physics of groups and low-mass clusters and the prospects for detailed
follow-up studies of these systems.}

\section{An introduction to the XMM-LSS survey}

The XMM-LSS survey (Pierre et al. 2004) is a medium-deep X-ray survey
undertaken to determine the values of key cosmological parameters on
the basis of the observed abundance and correlation statistics of
X-ray galaxy clusters. The XMM-LSS consists of a contiguous grid of
10-20~ks XMM pointings currently covering 5 deg$^2$. Observations
extending the X-ray coverage to 10 deg$^2$ are in progress during
2006. The XMM-LSS field is located at $\alpha = 02^{\rm h} 26^{\rm
m}$, $\delta = -5^\circ$ and overlaps with Canada France Hawaii Wide
Synoptic Survey (optical), the UKIDSS Deep Extragalactic Survey (NIR),
the SWIRE survey (mid-IR), the VIMOS-VLT Deep Survey (VVDS - optical
spectra) and is accompanied by additional imaging at radio and SZ
wavebands - see Figure 1 for more details.

\begin{figure}[t]
\centering
\psfig{figure=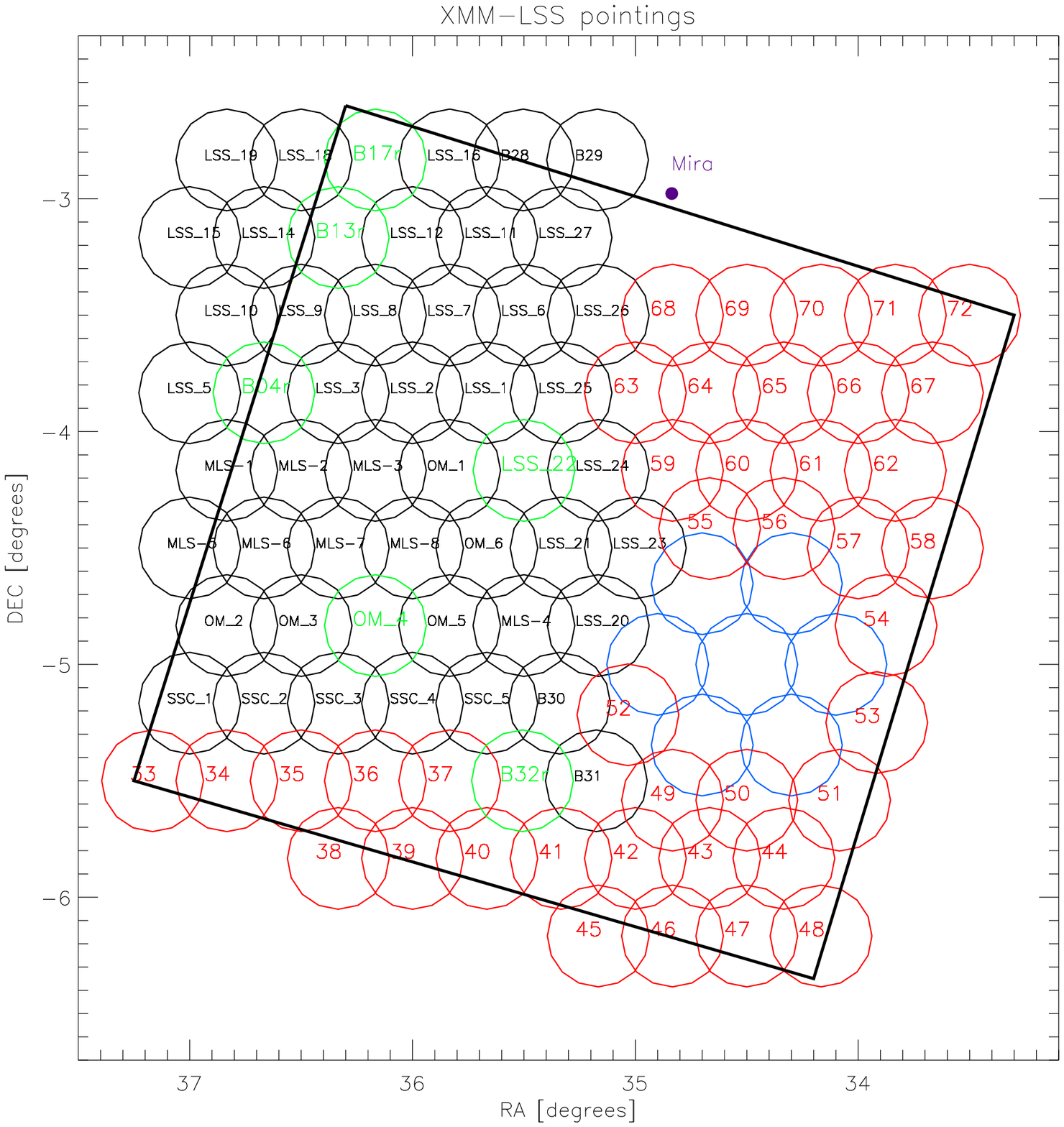,height=6.5cm}
\psfig{figure=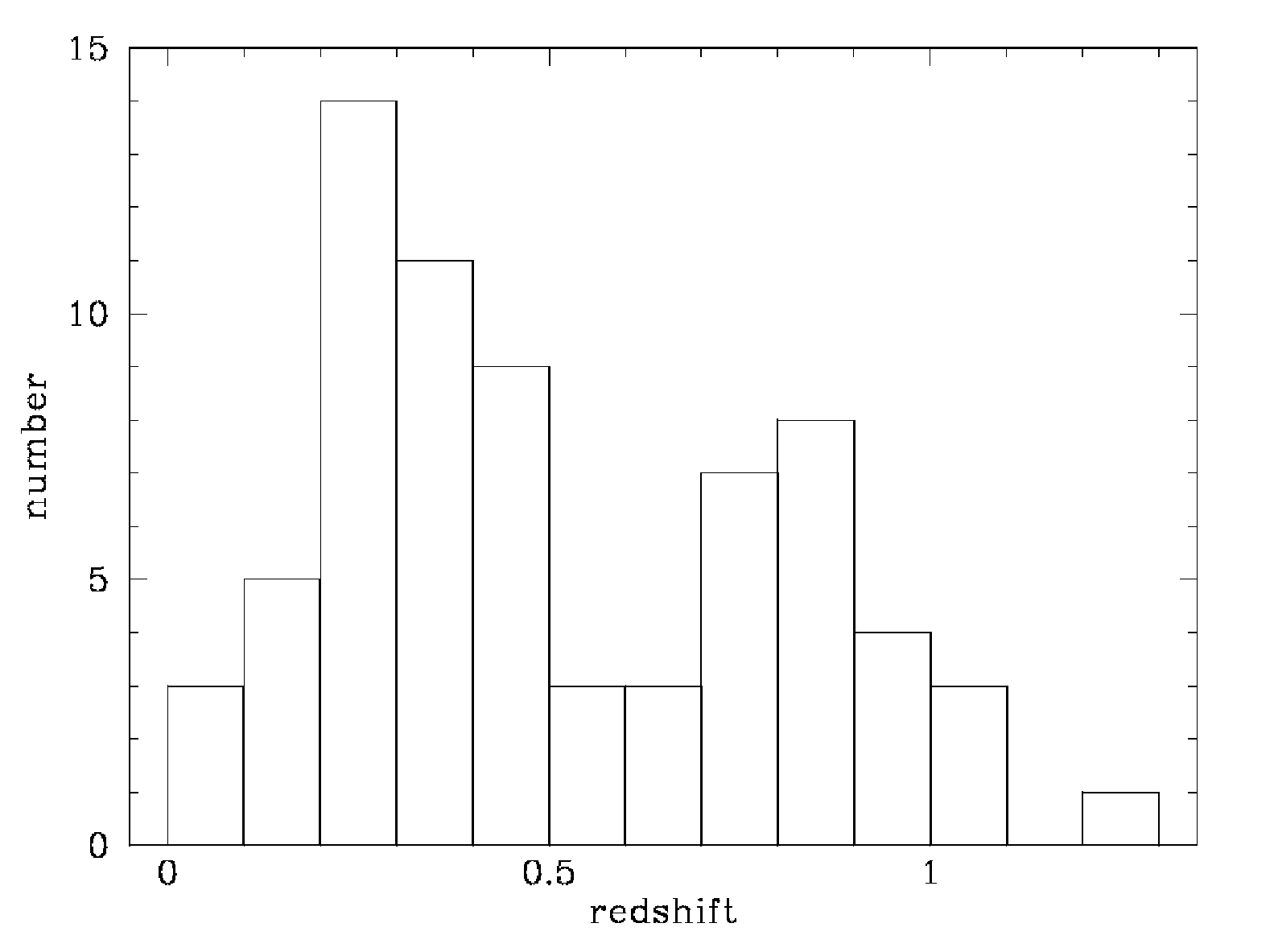,height=6.5cm}
\caption{Left panel: Field coverage of the XMM-LSS. Circles represent
individual XMM pointings (black$=$ pre-A05 pointings; red$=$A05
pointings; blue$=$Subaru-XMM deep survey: {\tt
www.naoj.org/Science/SubaruProject/SDS/}); green$=$pointings
compromised due to high background periods). The square indicates the
SWIRE field (Lonsdale et al. 2003). Right panel: redshift histogram of
all 71 clusters detected within XMM-LSS to date.}
\end{figure}

The XMM-LSS currently contains X-ray structures over the redshift
interval $0.05<z<1.22$ (Figure 1) yet is dominated numerically by
cooler, lower mass clusters $kT < 4$~keV at redshifts $0.2<z<0.6$. The
nominal flux limit of XMM-LSS is approximately $10^{-14}$ \eunit\ in
the [0.5-2] keV energy band and corresponds to a bolometric X-ray
luminosity $L_X \approx 10^{43}$ ergs \, s$^{-1}$ at
$z=0.6$\footnote{Assuming a surface brightness model characterised by
$\beta = 2/3$ and core radius $=180$~kpc.}. This sample represents a
considerable extension of the distance over which such systems can be
identified in moderate depth, moderate field X-ray surveys and it is
now possible to study the physical evolution of X-ray clusters
selected uniformly over both a wide interval of redshift and X-ray
luminosity. In particular, two broad areas of research that we will
touch upon in these proceedings are a) understanding the evolution of
the intracluster medium (ICM) through observed correlations between
X-ray luminosity, X-ray temperature and the cluster velocity
dispersion $\sigma_v$, and b) determining the extent to which
intermediate redshift groups and low mass clusters act as sites of
continuing galaxy evolution.

\section{Compiling the XMM-LSS sample}

\subsection{Creating a high-quality sample of X-ray clusters}

A detailed description of the procedures used by the XMM-LSS to
compile a highly complete sample of X-ray clusters displaying low or
no contamination is provided by Pacaud et al. (2006) and we summarize
the main considerations here. The raw X-ray observations were reduced
using the standard XMM Science Analysis Software (XMM SAS) and raw
photon images were produced by combining the data from the pn, MOS1
and MOS2 detectors to produce an image with a scale of 2.5 arcseconds
per pixel. Cluster detection was performed using a wavelet based
algorithm and was restricted to the inner 11 arcminutes of each XMM
pointing. Source classification was performed using a maximum
likelihood routine {\tt Xamin} (Pacaud et al. 2006) and employs the
parameters {\tt extent}, {\tt likelihood of extent} and {\tt
likelihood of detection}. In simple terms, X-ray source samples
generated using these parameters correspond to surface brightness
limited samples.  Table 1 describes the three classes of X-ray source
generated using this approach: C1, C2, and C3 (with C1 representing
the highest quality detections).

\begin{table}[t]
\caption{Properties of the three cluster classes generated by XMM-LSS}
\vspace{0.4cm}
\begin{center}
\begin{tabular}{|l|l|c|c|}
\hline
Class & Selection criteria & contamination & surface density \\
\hline
&&&\\
C1 & {\tt extent}$ > 5^{\prime\prime}$ && \\
& {\tt extent likelihood}$> 33$ & 0\% & $\sim5$ deg$^2$ \\
& {\tt detection likelihood}$> 32$ && \\
&&& \\
\hline
&&& \\
C2 & {\tt extent}$ > 5^{\prime\prime}$ && \\
& $15<${\tt extent likelihood}$< 33$ & 50\% & $\sim 3$ deg$^2$\\
& {\tt detection likelihood}$> 32$ && \\
&&& \\
\hline
&&& \\
C3 & $2^{\prime\prime}<${\tt extent}$ < 5^{\prime\prime}$ && \\
& {\tt extent likelihood}$> 4$ & unknown & $\sim 4$ deg$^2$\\
& $>30$ photons detected && \\
&&& \\
\hline
\end{tabular}
\end{center}
\end{table}

\subsection{Confirming optical redshifts}

The nature of each X-ray source \---\ whether a galaxy cluster at
specific redshift or a non-cluster source \---\ was determined via
optical photometry and spectroscopy. A combination of either
CTIO/MOSAIC $Rz^\prime$ (Andreon et al. 2004) or
CFHT/MEGACAM\footnote{Data are taken from the CFHT Wide Synoptic
Legacy Survey. See the URL {\tt www.cfht.hawaii.edu/Science/CFHTLS/}
for further details.} $ugriz$ imaging was used to associate the
location of each X-ray source with the spatial barycentre of a
significant overdensity of galaxies displaying characteristically red
colours.  Galaxies lying within a given colour tolerance of this ``red
sequence'' were flagged as candidate cluster members and given a high
priority in subsequent multi-object spectroscopic observations.  A
small number of low X-ray temperature ($\sim 1$~keV) groups at
moderate redshift ($z>0.2$) and local ($z<0.2$) optically poor groups
were not associated with a statistically significant galaxy
overdensity. These systems were inspected visually and spectroscopic
targets were assigned manually. Spectroscopic data were reduced using
standard procedures described in detail in previously published
XMM-LSS samples (Valtchanov et al. 2004; Willis et al. 2005; Pierre et
al. 2006). Cluster redshift values were then computed as the
unweighted mean of all galaxies within $\Delta z = 0.02$ of the
visually assigned redshift peak.

\section{The Luminosity-Temperature relation}

\subsection{Fitting robust temperatures to faint, low-temperature clusters}

Galaxy clusters selected within the C1 and C2 classes display total
X-ray count levels as low as 100 photons. However, by virtue of the
coincidence of temperature sensitive Iron L-shell emission at
energies of order $kT<4$~keV with the energy range of peak sensitivity
of the XMM detectors, it remains possible to fit reliable temperature
values to these faint systems. The details of the spectral analysis
procedure are described in Willis et al. (2005) and fit an absorbed
APEC hot plasma model (Smith et al. 2001) to the source spectral
energy distribution (SED) using the {\tt Xspec} routine (Arnaud
1996). The main modification is that the source and background spectra
are rebinned to a common spectral scale such that the background SED
displays a minimum of 5 counts per spectral bin. Extensive simulations
indicate that this procedure generates reliable temperature estimates
(i.e. a small systematic offset from the true temperature and {\tt
Xspec} errors that provide a realistic measure of the distribution of
fitted temperatures) when fitting clusters as cool as $kT=1$~keV with
as few as 100 total counts.

\subsection{The $L-T$ relation}

Figure 2 displays the X-ray luminosity temperature relation for 12
groups and clusters analysed within an initial XMM LSS sample (Willis
et al. 2005). This sample represents a mix of seven C1$+$C2 and five
C3 systems. Bolometric luminosity values are computed by extrapolating
the measured flux in the [0.5-2]~keV energy band to all wavelengths
employing the fitted spectral emission model. In addition, each X-ray
system is fitted with a spatial emission model consisting of a
circular $\beta$-model convolved with the local PSF. The value of
$r_{500}$ for each system is determined using the fitted temperature
and the mass-temperature relation of Finoguenov, Reiprich \&\
Bohringer (2001) and the spatial emission model is extrapolated to
this scale radius to present brightness measures as $L_{bol}(r_{500})$

\begin{figure}[h]
\centering
\psfig{figure=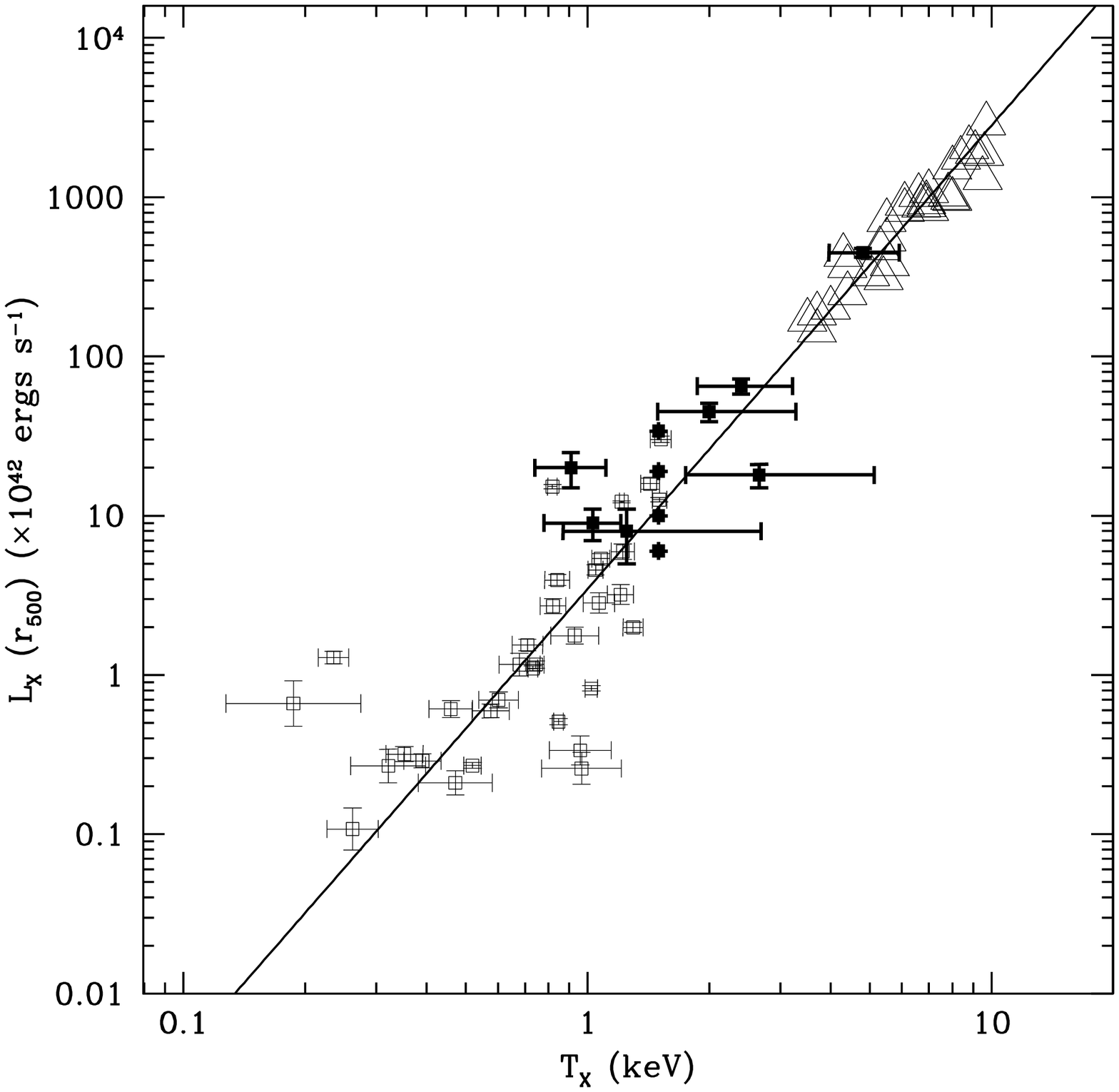,height=7.0cm}
\psfig{figure=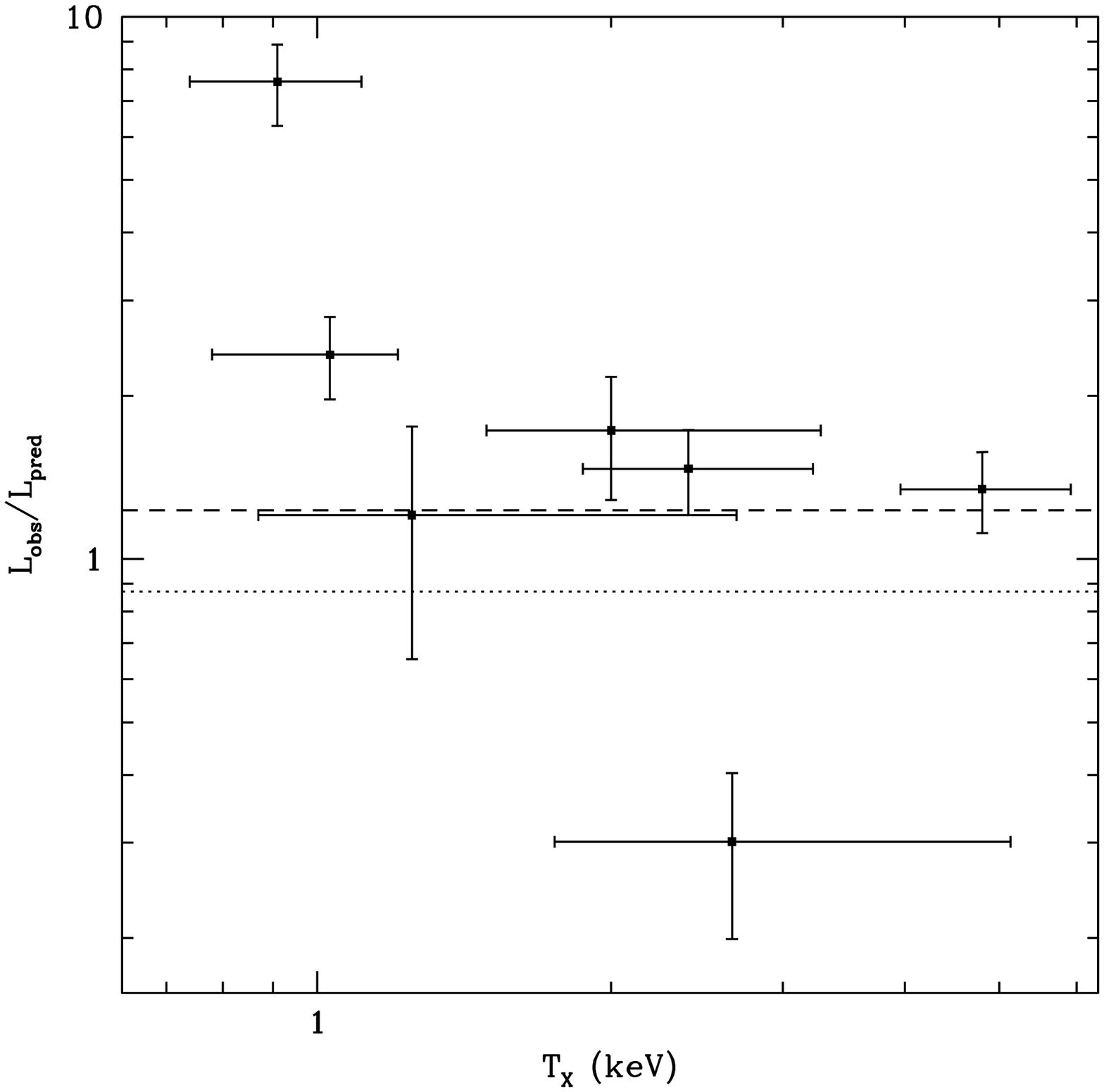,height=7.0cm}
\caption{Left panel: Distribution of X--ray luminosity computed within
a scale radius $r_{500}$ and temperature for XMM--LSS groups and
clusters presented in Willis et al. 2005 (solid squares). Also
indicated are values of X--ray luminosity and temperature determined
for the low redshift group sample of Osmond and Ponman (2004) (open
squares) and for the cluster sample of Markevitch (1998) (open
triangles). The solid line indicates an orthogonal regression fit to
the $L_X$ versus $T_X$ relation for both the group and cluster sample
incorporating a treatment of the selection effects present in each
sample ($\log L_X = 2.91 \log T_X + 42.54$). Right panel: Enhancement
factor, $F = L_{obs}/L_{pred}$, computed for six XMM--LSS groups and
clusters located at $z \le 0.6$ plotted versus the X--ray temperature
of each system (see Willis et al. 2005 for additional
details). Horizontal lines indicate expected values of $F$: the short
dashed line indicates the value $F=1.23$ expected from self--similar
considerations. The dotted line indicates the value of $F$ expected at
$z=0.4$ based upon Ettori et al. (2004).}
\end{figure}

Examination of the $L-T$ relation for this initial XMM-LSS sample
indicates that a small amount of positive luminosity evolution is
present compared to a local $L-T$ relation (see Figure 2 for more
details). At a given temperature, these systems are in the median 1.46
times more luminous compared to a local $L-T$ relation, whereas,
applying self-similar scaling one would expect a luminosity
enhancement of order 1.23, i.e. positive luminosity evolution is
present at the level of 20\%. Given the modest size and the
statistically incomplete nature of this initial sample, we regard
these results as in need of confirmation.  Ettori et al. (2004) report
evolution weaker than the self--similar expectation (i.e. slightly
negative luminosity evolution) from a sample of 28 clusters at $z>0.4$
with gas temperatures $3~{\rm keV} < kT < 11~{\rm keV}$. The combined
effect of self--similar scaling with the negative evolution reported
by Ettori et al. (2004) would result in an enhancement factor $F=0.86$
at a $z=0.4$ (see Figure 2). Though the overlap of the Ettori et
al. (2004) sample and the systems contained in the present work is
limited, further work is clearly required to understand the $L-T(z)$
evolution of low temperature systems.

In particular, XMM-LSS is currently generating a large sample of
galaxy groups and clusters displaying $kT<4$~keV and $0.2<z<0.6$. More
importantly however, the X-ray selection function has been computed as
a function of apparent brightness (count rate or flux) and core radius
of the fitted surface brightness profile (Pacaud et al. 2006). In
addition, the C1 sample is temperature complete \---\ every C1 cluster
has a fitted temperature. Equipped with a large sample and an explicit
selection function, the XMM-LSS is currently well placed to perform a
robust investigation of the $L-T$ evolution of X-ray groups and
clusters.

\section{A detailed spectroscopic look at three low-T systems}

Low temperature X-ray groups and clusters at $z>0.2$ are being
revealed in large numbers by XMM-LSS for the first time. A detailed
magnitude limited spectroscopic study of a subset of these systems is
clearly required to a) determine their dynamical state and search for
evidence of recent mergers (and to assess their effect upon the ICM)
and b) perform an unbiased (or ``least biased'') survey of the member
galaxies in an attempt to correlate the extent of recent star
formation with merging activity and the properties of the X-ray
emitting gas (via deviations from the $L-T$ relation).

\begin{figure}[t]
\centering
\psfig{figure=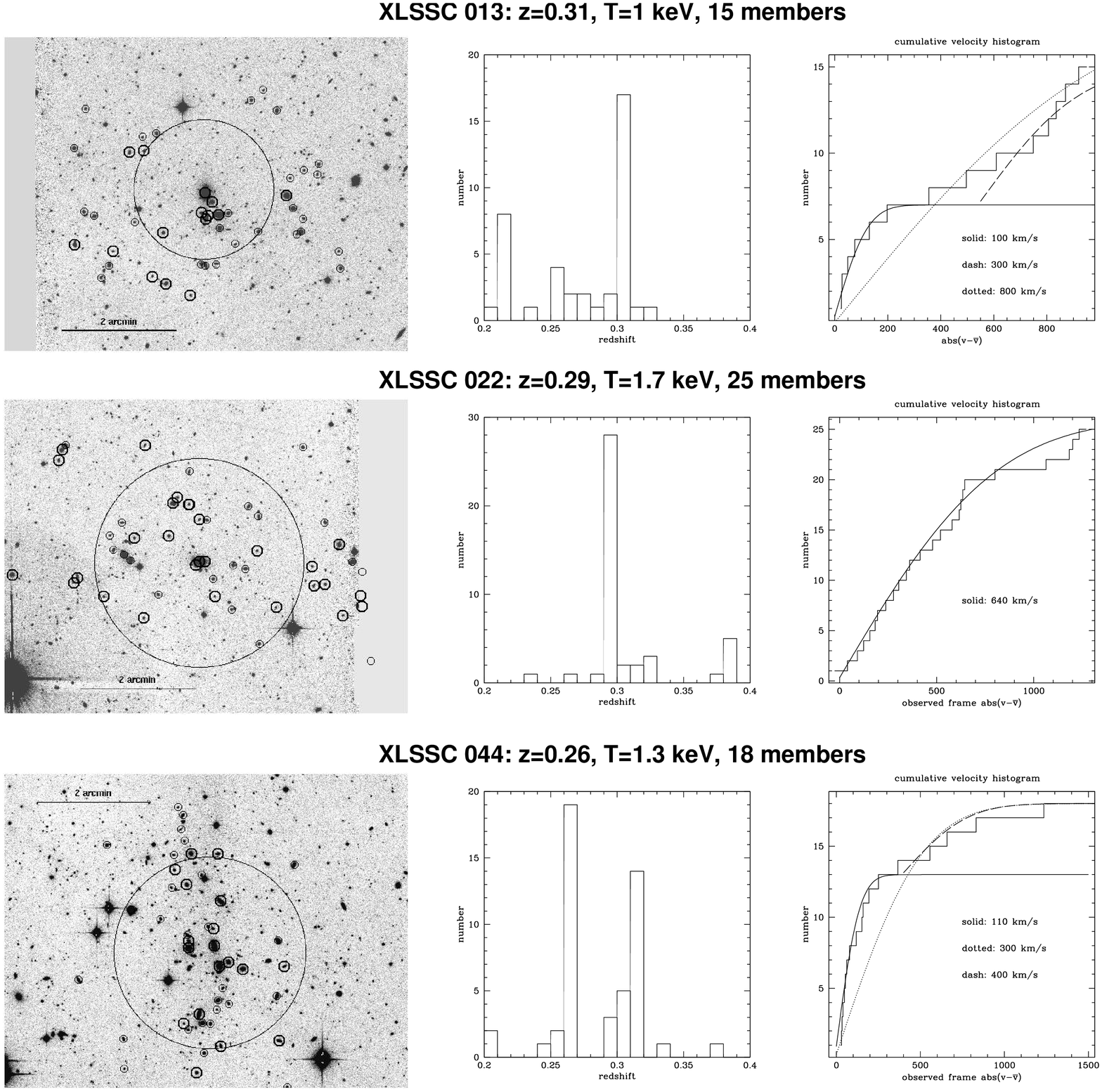,height=15.5cm}
\caption{Three low temperature X-ray groups at $z\sim0.3$ observed
with highly complete spectroscopy to $R<21.5$. Left panels: CFHT
Legacy Survey Wide Synoptic $i$-band data for each X-ray group. North
is up and East is left. The overplotted small circles indicate
galaxies confirmed at the cluster redshift (heavy circle) and
unassociated field galaxies (light circles). The large circle
indicates the radius $r_{500}$ for each group. Centre panels: redshift
histogram of each field over $0.2<z<0.4$. XLSSC 044 in particular
displays a number of significant redshift peaks along the
line-of-sight. The association of the $z=0.26$ galaxies with the X-ray
emission is clear but additional structure may affect the derived
X-ray luminosity and temperature. Right panels: the cumulative
distribution of observed frame member velocities relative to the
systemic velocity. Overplotted are sample Gaussian distributions
describing different velocity dispersions. A comparison of the data to
these simple models indicates that groups XLSSC 013 and 044 appear to
be composed of distinct velocity components whereas XLSSC 022 appears
to be a very regular system.}
\end{figure}

During ESO Period 76 we obtained spectroscopy of all galaxies
satisfying $R<21.5$ in the fields of three $kT=1-2$~keV groups at
$z\sim 0.3$ (see Figure 3). The three systems XLSSC 013, 022 and 044
are described in detail in Pierre et al. (2006). Two of the three
systems, XLSSC 013 and 044, display evidence both for velocity
substructure and for additional projected structures along the
line-of-sight. These X-ray groups display velocity distributions that
are inconsistent with simple Gaussian distributions. In general, each
system appears to display at least one other velocity component offset
from the systemic velocity. The possibility that we are observing
merging systems and (in the case of 022) a filament in projection,
have important consequences for how we interpret such systems as
reliable cosmological probes, e.g. both effects will affect the slope
and scatter of the mass-temperature relation. In contrast, the group
XLSSC 022 (the hottest of the three) displays a very regular velocity
structure. We aim to investigate these systems further in ESO Period
79 with VLT/FORS2 spectroscopy to a deeper magnitude limit.

\section{Future work}

Clearly there is much that remains to be understood regarding X-ray
groups and low mass clusters at intermediate redshifts. Of principle
important is an understanding of the distribution of low temperature
systems on the mass--temperature relation exhibited by hotter, more
massive systems. A relatively detailed knowledge of the
mass--temperature relation is required to compare observed correlation
and abundance statistics to the predictions of dark matter clustering
in a $\Lambda$CDM universe. To this end, we are continuing our
detailed investigation of the dynamical state of X-ray groups at
intermediate redshift with the dual aim of determining dynamical
masses for these systems and understanding the extent to which their
recent evolution is driven by merging. Combined with a detailed
understanding of dynamical effects in individual groups and clusters,
we are continuing to investigate the evolution of the $L-T$ relation
for the XMM-LSS sample, incorporating a detailed treatment of the
X-ray selection function (Pacaud et al. 2007 in prep.). In tandem with
existing X-ray studies of individual groups and clusters at
intermediate redshift, XMM-LSS is therefore providing an important
census of a poorly understood and potentially very active regime of
structure formation.

\section*{Acknowledgments}
JPW gratefully acknowledges the generous financial support of the
conference organisers in attending the 26th Moriond Astrophysics
Meeting.

\end{document}